\documentclass[sn-basic]{sn-jnl}

\usepackage{graphicx}%
\usepackage{multirow}%
\usepackage{amsmath,amssymb,amsfonts}%
\usepackage{amsthm}%
\usepackage{mathrsfs}%
\usepackage[title]{appendix}%
\usepackage{xcolor}%
\usepackage{textcomp}%
\usepackage{manyfoot}%
\usepackage{booktabs}%
\usepackage{algorithm}%
\usepackage{algorithmicx}%
\usepackage{algpseudocode}%
\usepackage{listings}%
\usepackage{subfig}
\usepackage{hhline}
\usepackage[markup=underlined]{changes}
\usepackage[
  automark,
  autooneside=false,
  headsepline
]{scrlayer-scrpage}
\clearpairofpagestyles
\ihead{\scriptsize Accepted to Automated Software Engineering Journal}
\ohead{\scriptsize Chia-Yi Su and Collin McMillan}

\theoremstyle{thmstyleone}%
\theoremstyle{thmstyletwo}%

\theoremstyle{thmstylethree}%

\begin{document}

\title[Article Title]{Distilled GPT for Source Code Summarization}


\author*[1]{\fnm{Chia-Yi} \sur{Su}}\email{csu3@nd.edu}

\author[1]{\fnm{Collin} \sur{McMillan}}\email{cmc@nd.edu}

\affil*[1]{\orgdiv{Department of Computer Science}, \orgname{University of Notre Dame}, \orgaddress{\street{Holy~Cross Dr}, \city{Notre Dame}, \postcode{46556}, \state{Indiana}, \country{USA}}}




\abstract{A code summary is a brief natural language description of source code.  Summaries are usually only a single sentence long, and yet form the backbone of developer documentation.  A short descriptions such as ``changes all visible polygons to the color blue'' can give a programmer a high-level idea of what code does without the effort of reading the code itself.  Recently, products based on Large Language Models such as ChatGPT have demonstrated a strong ability to write these descriptions automatically.  However, to use these tools, programmers must send their code to untrusted third parties for processing (e.g., via an API call).  This loss of custody is not acceptable to many organizations.  In this paper, we present an alternative: we train an open source model using sample output generated by GPT-3.5 in a process related to knowledge distillation.  Our model is small enough (350m parameters) to be run on a single 16gb GPU, yet we show in our evaluation that it is large enough to mimic GPT-3.5 on this task.}




\maketitle

\section{Introduction}
\label{intro}

A code summary is a brief, natural language description of source code.  Summaries are typically only a single sentence.  When reading a Java method, for instance, a programmer may start with the Javadoc sentence ``changes all visible polygons to the color blue.''  The summary provides a quick way for the programmer to understand what the method does without having to read the code itself.  The benefits of summaries in documentation have been studied for decades~\citep{haiduc2010use}, and Software Engineering (SE) research has long sought to automate the process of writing them, to reduce manual effort by programmers, support under-documented legacy programs, and build accessibility tools~\citep{robillard2017demand}.  Code summaries form the backbone of much documentation for programmers, and the dream of automatic generation of these summaries has been described as a ``holy grail'' of SE research~\citep{allamanis2018survey, forward2002relevance, leclair2019neural}.

Recently, the dream seems within reach.  Years of effort on neural code summarization techniques has culminated in products such as Copilot~\citep{github2022copilot} and ChatGPT~\citep{openai2022chatgpt}, which exhibit an ability to describe arbitrary code~\citep{ma2023scope}.  At the heart of these products is a language model that is trained using big data input.  The language model in the most powerful products may be tens or hundreds of billions of parameters, and the data input often includes trillions of tokens, such as the entirety of public GitHub repositories, plus StackOverflow, Wikipedia, etc.  The effectiveness of these products has captured the public imagination and helped drive a new wave of research~\citep{sun2023automatic}.  Like in many research areas, the decades-long effort toward automatic code summarization suddenly seems at hand.

Yet a major problem looms.  For programmers to use these tools, they must send their code to third parties for processing.  An IDE plugin wishing to use GPT-3.5, for example, must harvest code from the programmer's codebase and send it via an API call to OpenAI.  This call is a loss of data custody and a non-starter for many institutions~\citep{derner2023beyond}.  In addition, the closed nature of these products has caused controversy among researchers, who point out a lack of reproducibility, potential data contamination from public test sets to private training data, and resultant loss of scientific rigor~\citep{hellendoorn2021growing}.  The situation for many programmers is that the technology to automate a major portion of code summarization exists, but it is not usable.

In this paper, we present an alternative: knowledge distillation from a large model (GPT-3.5) to smaller models.  Our paper has three key novel research contributions: 

\begin{enumerate}
\item We present a study comparing summaries from GPT-3.5 to the reference summaries written by human programmers, and show that the generated summaries tend to be superior, indicating they are good source of training data. (Section~\ref{sec:studyone})
\vspace{1mm}
\item We present a study of knowledge distillation of these summaries for the size of model (38m - 15.5B parameters) and size of training data (170k - 2.15m samples).  We collect 2.15m summaries generated by GPT-3.5 for Java methods.  We use a simple prompt and methods from open-source Java programs.  As foundation models, we compare \texttt{jam}~\citep{su2023language} and \texttt{starcoder}~\citep{li2023starcoder}.  The \texttt{jam} model is pretrained on 52m Java methods and has an easily-searchable dataset to ensure reproducibility and a controlled experimental framework.  The \texttt{starcoder} model is much larger and has a larger pretraining dataset, but is also more expensive and has more non-controllable experimental variables due the the dataset. (Section~\ref{sec:studytwo})
\vspace{1mm}
\item We evaluate our distilled model against GPT-3.5 in a study with human experts. (Section~\ref{sec:studythree})
\end{enumerate}

We release all code and implementation details.  The model we recommend from our experiments can be run from a workstation with a single 16GB GPU, which is relatively low cost for many organizations.  This low cost and open source structure enables programmers to access automatic code summarization while keeping data custody. Although there are some papers that have already formulated the code summarization as a fine-tuning problem  such as~\citet{wang2021codet5, bender2021danger} and studied knowledge distillation for smaller models from larger models~\citet{hsieh2023distilling, yu2023large}, we thoroughly explore the data and model size for knowledge distillation on code summarization and conduct the human study to compare the language models generated summary and human reference with human experts.

\section{Background and Related Work}
\label{sec:background}

This section discusses the two key areas of background and related work: code summarization and knowledge distillation.

\begin{table*}[!b]
    \centering
	{\small
    
		\begin{tabular}{p{4cm}p{0.4cm}p{0.4cm}p{0.4cm}p{0.4cm}p{0.4cm}}
        \centering
			& I     & N          & G          & T & C                \\
			\citet{mcburney2016automated}						& x &  	&   &   & x \\
   \citet{iyer2016summarizing}						&  &  x	&   &   &  \\
			\citet{rodeghero2017detecting}		& x &  	&   &   & x \\
			\citet{fowkes2017autofolding}			& x &  	&   &   &   \\
   \citet{loyola2017neural}			&  & x 	&   &   &   \\
    
			\citet{jiang2017automatically}			&   & x	&   &   &   \\
   \citet{hu2018deep}							&   & x	& x &   &   \\
   \citet{hu2018summarizing}							&   & x	&  &   &   \\
			
			\citet{allamanis2018learning}			&   & x	& x &   &   \\
   \citet{wan2018improving}							&   & x	&  &   &   \\
   \citet{liang2018automatic}							&   & x	&  &   &   \\
			\citet{alon2019code2seq, alon2019code2vec}	&   & x	& x &   &   \\
			\cite{gao2019neural}						&   & x	&   &   &   \\
			\citet{leclair2019neural}				&   & x	& x &   &   \\
			\citet{nie2019framework}					&   & x	&   &   &   \\
   \citet{lu2019learning}			&  & x 	&   &   &   \\
   \citet{gao2019neural}			&  & x 	&   &   &   \\
			\citet{haque2020improved}					&   & x	&   &   & x \\ 
   \citet{haldar2020multi}			&  & x 	&   &   &   \\
			\citet{zugner2021languageagnostic}	&   & x	& x &   &   \\
			\citet{liu2021retrievalaugmented}			&   & x	& x &   &   \\
			\citet{bansal2021project}				&   &x       &    &   & x \\
   \citet{wang2021codet5}			&   &x      &    &   & \\
   \citet{bender2021danger}				&   &x       &    &   & \\
		\end{tabular}
	}
	\vspace{0.2cm}
	\caption{Snapshot of the past five years in source code summarization.  Column $I$ stands for IR-based techniques.  $N$ means neural network-based.  $G$ means the code is modeled as a graph.  $T$ means Transformer designs.  $C$ means learning from context.}
	\label{tab:screlated}
\end{table*}

\subsection{Source Code Summarization}

The term ``source code summarization'' was coined in 2010 by~\citet{haiduc2010use} to refer to the task of writing natural language descriptions of code.  Until 2017, most of research methods focused on IR-based and template-based methods. From 2017 to present, neural models for code summarization becomes the most dominant research direction. Table~\ref{tab:screlated} shows the history and families of different methods. Although the most dominant research line is neural models, there are different families of neural models, which include better modeling of code itself and using more context. For example,~\citet{leclair2019neural, alon2019code2seq, alon2019code2vec} combined AST with source code and  \citet{allamanis2018learning} modeled AST as a graph for neural models.~\citet{haque2020improved} applied the attention mechanism to the file context.~\citet{bansal2021project} combined information different software projects.

More recently,~\citet{wang2021codet5, bender2021danger} introduce the technique to fine-tune Large Language Models (LLM) for code summarization. Although some proprietary LLM such as OpenAI's ChatGPT has demonstrated the great capability on program comprehension, we cannot avoid data leak problems because of the inaccessibility of training data on these models. In this paper,  we examine the data and model size for fine-tuning and distill the knowledge from GPT-3.5 by training small models with public and controllable datasets and mimic GPT-3.5's capability for code summarization.

\subsection{Knowledge Distillation}

Knowledge Distillation refers to teaching a small machine learning model to behave like a larger one for a niche task~\citep{hsieh2023distilling, yu2023large, wang2021knowledge}.  Knowledge distillation is useful in scenarios where a large model may be capable of many tasks or even considered general purpose, such as ChatGPT or Copilot, but is too expensive or impractical to use for certain specific tasks.  A classic application is in image classification, where a powerful model capable of classifying many types may be used to teach a smaller model with some tradeoffs, such as lower accuracy or recognizing fewer categories~\citep{gou2021knowledge, wang2021knowledge, zagoruyko2016paying}.  More recently, general purpose models such as GPT-3.5 have been used to teach smaller models to perform question-answering tasks~\citep{zhang7llama}, assessment of student answers~\citep{li2023distilling}, follow specific types of instructions~\citep{tang2023domain}, and to improve output from existing smaller chatbots~\citep{chen2023phoenix}.

An important point in knowledge distillation is that there is almost always a trade made in exchange for the reducing model size.~\citet{gudibande2023false} make the point that small models attempting to replicate all capabilities of ChatGPT, for instance, are likely to face problems with generalization to prompts unlike those in the training samples.  They highlight how the benefit of knowledge distillation is focused on training small models to perform specific tasks -- one should not necessarily expect the small model to do all tasks well, but the small model can mimic the large one by specializing on a single task.  In this paper, we focus on the task of code summarization, and demonstrate how a small model can perform on par with the large one in this specialty.

\section{GPT-3.5 and Human-written References}
\label{sec:studyone}

This section discusses our comparison of source code summaries generated by GPT-3.5 to summaries written by human programmers. We also compare the summaries generated by the model trained with GPT-3.5 summaries and the summaries written by human programmers. The purpose of this comparison is to determine whether summaries generated by GPT-3.5 are suitable replacements for human-written ones.  This comparison is necessary in this paper because we seek to distill GPT-3.5's ability to generate summaries for a smaller model, so we should measure the quality of the summaries GPT-3.5 generates.  If GPT-3.5 generates poor quality summaries, then it would not be suitable for distillation.  Thus, we ask the Research Question (RQ):

\begin{description}
\item \textbf{RQ1} How well do summaries generated by GPT-3.5 compare to human-written reference summaries, across key quality criteria established in relevant literature?
\end{description}
By ``human-written reference summaries'', we mean code summaries written by the programmers or other team members who wrote the underlying software, such as the summary sentence of the Javadocs for Java methods.  By ``key quality criteria'', we mean the concepts of Accuracy, Completeness, and Concision that were first used to evaluate code summaries over ten years ago by~\cite{sridhara2010towards} and have been used in numerous studies since~\cite{mcburney2016automated}.  We will establish how we measure these concepts in the next subsection.  Note we also measure overall preference of one summary to another, for a gauge on how people balance the quality criteria.

\subsection{Research Method}
\label{sec:rq1method}

Our research method is a survey in which we show programmers different code and summaries of that code, and ask them to rank the summaries in four questions.  We designed our survey to be consistent with years of best-practice in evaluating summaries, namely from~\cite{sridhara2010towards, mcburney2016automated, bansal2021neural}, including the wording we use in the survey questions.  The survey has four questions per summary, divided over two pages.  On the first page, the survey shows a Java method and a summary of that method, plus these three questions:

\begin{enumerate}
    \item Independent of other factors, I feel that the summary is accurate.
    \item The summary is missing important information, and that can hinder the understanding of the method.
    \item The summary contains a lot of unnecessary information.
\end{enumerate}

The first question is intended to measure accuracy, the second measures completeness, and the third measures concision.  Following each question are four radio buttons to select one of: ``Strongly Agree'', ``Agree'', ``Disagree'', and ``Strongly Disagree.''  On the next page, the survey shows the same Java method and summary, plus another summary for the same Java method.  The first summary is either the summary from GPT-3.5 or the human-written reference (decided randomly).  The second summary is the alternative.  The second page also shows a single question:

\begin{enumerate}
    \setcounter{enumi}{3}
    \item Overall, which summary is better in your opinion?
\end{enumerate}


\begin{figure}[!h]
    \centering
    \subfloat[\centering Page One]{{\includegraphics[width=6cm, height=4cm]{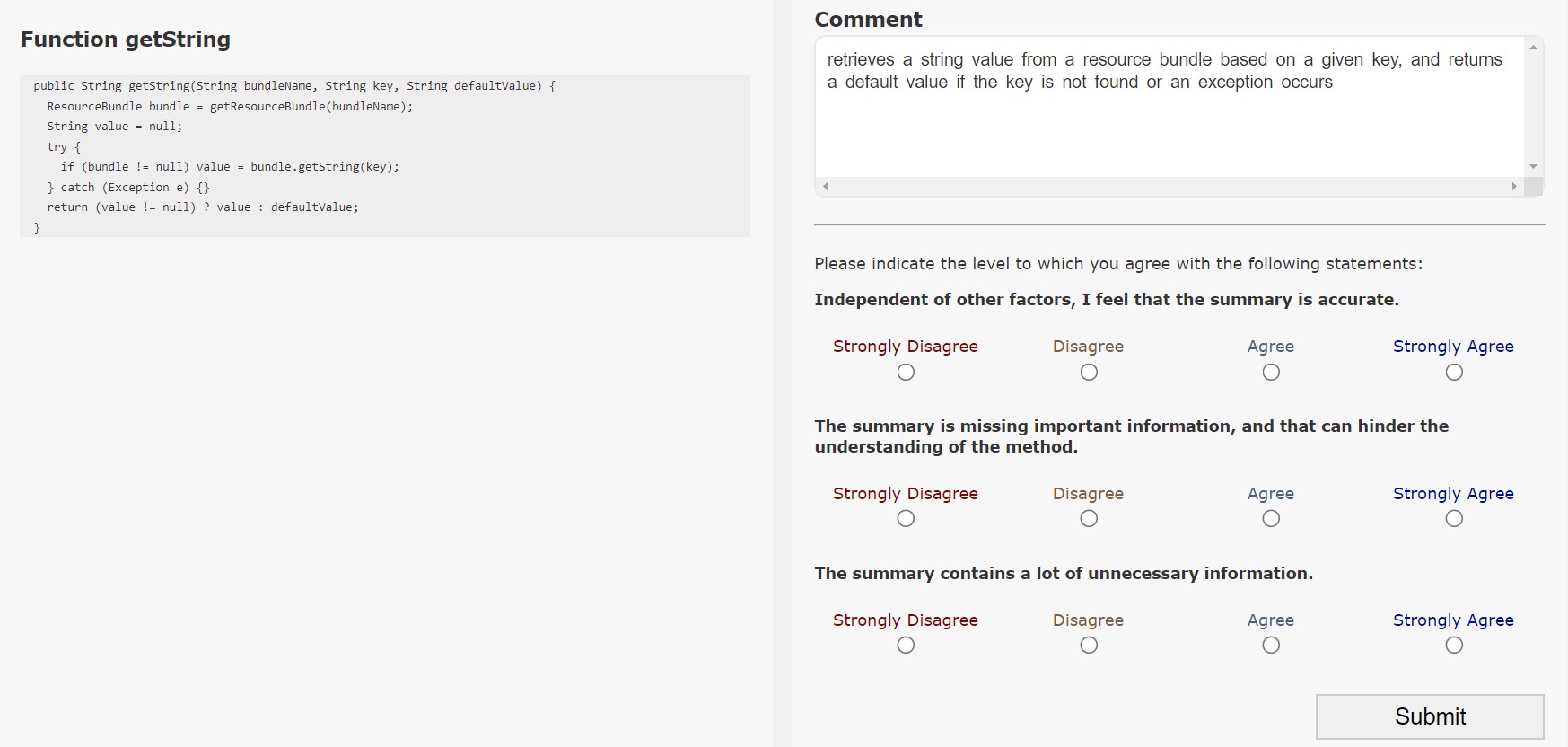} }}
    \qquad
    \subfloat[\centering Page Two]{{\includegraphics[width=6cm, height=4cm]{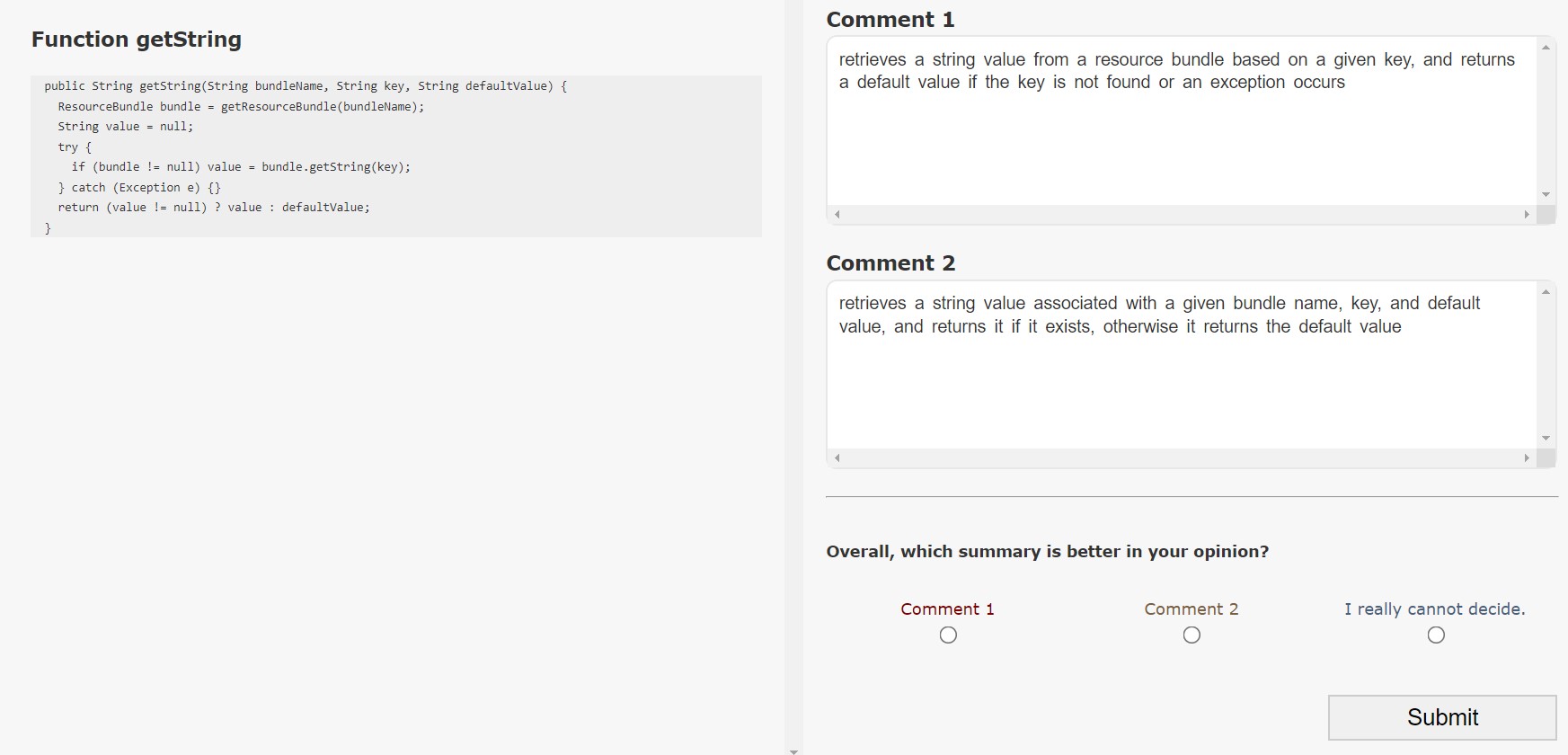} }}
    \vspace{1mm}
    \caption{Example of the two pages of our survey for each Java method.}
    \label{fig:surveypages}
\end{figure}


Following this question are three options: ``Summary 1'', ``Summary 2'', and ``I really cannot decide.''  We phrased the third option such that participants would be encouraged to select between the two summaries, but still have an option to avoid the question in very difficult or impossible cases, such as if the summary were the same or both were illegible.  For quality control (see Section~\ref{sec:threatsone}), we also asked participants to enter a rationale for their answer to this question.  Figure~\ref{fig:surveypages} shows each page.

\subsection{Subject Source Code, Summaries, Participants}

We obtained the subject source code from the \texttt{funcom-java-long} dataset provided by~\citet{su2023language}.  This dataset is a revision of the dataset provided by~\citet{leclair2019recommendations} to include various fixes such as those proposed by~\citet{bansal2021project, shi2022are}.  This dataset includes a training set of around 170k Java methods paired with summaries written by human experts (the origin of these summaries was Javadocs provided with the source code), as well as a test set of around 8k Java methods.  The dataset also includes a total of over 52m Java methods that do not contain human-written summaries.  The \texttt{funcom-java-long} dataset is diverse in that it contains over 50k Java projects from many domains collected over at least one decade.  The test set of 8k methods represents 880 projects.

To obtain summaries written by GPT-3.5, we used a prompt in the format:

\vspace{1mm}

\begin{verbatim}
    Write a one sentence description of this Java method:
\end{verbatim}

\vspace{1mm}

Followed by the source code for the method.  We collected summaries from GPT-3.5 using this prompt for a total of 2.15m Java methods.  The 2.15m Java methods included all 170k from the \texttt{funcom-java-long} training set, all 8k from the \texttt{funcom-java-long} test set, plus 2m additional methods randomly selected from the 52m Java methods in the dataset that do not have human-written summaries.  We filtered approximately 20k summaries which were either empty or not in English.

The survey randomly selected 30 Java methods from the 8k test set to show to each participant.  We chose 30 because we found that participants tended to spend between two and three minutes per method, and we aimed to keep the survey time to a maximum of 90 minutes to prevent fatigue bias~\citep{sievertsen2016cognitive}.

We recruited 15 participants for each study via the Prolific platform\footnote{\url{https://www.prolific.co/}}.  We used Prolific's features to filter for people who were at least 25 years of age, were located in the United States or United Kingdom, and had a university degree in Computer Science or Computer Engineering.  We describe additional filters for biases in the Threats to Validity, Section~\ref{sec:threatsone}.  With 15 participants and 30 methods each, we collected feedback for 450 Java methods.  Approximately half showed the human-written summaries on the first survey page (therefore answering questions 1-3), with the remainder showing the GPT-3.5 summaries on the first page.  All saw both summaries on the second page. To reduce the subjective bias, we randomly selected 150 functions from the original 450 samples and we recruited additional five different participants (30 methods for each participant) to evaluate the methods with the same criteria and the same website via the Prolific platform.

\begin{table}[!b]
\begin{tabular}{p{4cm}p{8.5cm}}
\multicolumn{2}{l}{\begin{tabular}[c]{@{}l@{}}
private Time simulate() throws SimulationException\{
\\ \quad Time current = startSimulation();
\\ \quad try \{
\\ \qquad if (endTime.isLT(current)) \{
\\ \quad \qquad throw new SimulationException(""Requested time "" + endTime
\\ \qquad \qquad+ "" is smaller than current time "" + current + ""!"");
\\ \qquad while (isRunning()) \{
\\ \quad \qquad if (current.isGE(endTime)) \{ break; \}
\\ \quad \qquad current = continueSimulation(current, endTime);
\\ \quad \qquad Thread.yield();
\\ \quad \} finally \{
\\  \qquad finishSimulation();
\\ \quad \}
\\ \quad return currentTime();
\\ \}
\\ ~\end{tabular}} \\
GPT-3.5  &  simulates a process until a given end time and returns the final current time, handling exceptions if the end time is reached before the current time
                                      \\
Human Reference                                      & increase the simulation time and execute all events with an earlier time\\
Trained with GPT-3.5  &  simulates a simulation by checking if the requested time is greater than the current time and returns the simulation time, or throws a SimulationException if the requested time is not found \\
Trained with Human Reference                                      & simulate the simulation 
                                    
\end{tabular}

\vspace{3mm}
\begin{tabular}{p{4cm}p{8.5cm}}
\multicolumn{2}{l}{\begin{tabular}[c]{@{}l@{}}
private Literal promoteDecimal(Literal numeral)  \{
\\ \quad Long result;
\\ \quad try \{
\\ \qquad result = Long.valueOf(numeral.getLabel());
\\ \quad \} catch (NumberFormatException e) \{
\\  \quad throw new TypeError(""Cannot promote non-numeral to a decimal value"");
\\  \quad \}
\\ return this.factory.createLiteral(result.toString(), SPARQLConstants.DECIMAL\_TYPE); \\ \}\\ ~\end{tabular}} \\
GPT-3.5                                              & promotes a Literal numeral to a decimal value by converting it to a Long and creating a new Literal with the converted value and the DECIMAL\_TYPE
                                      \\
Human Reference                                      & promotes a literal to a decimal datatype reparsing the label \\
 Trained with GPT-3.5                                              & promotes a given Literal to a decimal type and returns the promoted Literal \\
 Trained with Human Reference                                      & promotes a literal to a decimal literal                                    
\end{tabular}


\caption{Examples of summaries generated by GPT-3.5 and the human reference for two Java methods from the dataset.}
\label{tab:sumexamples}

\end{table}

\subsection{Threats to Validity}
\label{sec:threatsone}

The key threats to validity in this study are: 1) the participants, 2) the GPT-3.5 version and prompt, and 3) the subject Java methods.  The participant pool can be a threat to validity because online survey participants can fake work history.~\citet{danilova2021you} recommend programming-based screening questions, but~\citet{ghorbani2023autonomy} point out that these questions are now easily circumvented with online AI-based tools such as Copilot and ChatGPT.  Therefore, we manually inspect each participant's survey results for clear patterns of fraud.  We rejected one participant who completed the survey in under fifteen minutes (thirty seconds per method). 

The GPT-3.5 version and prompt are threats to validity because GPT-3.5 is a commercial product and subject to change without notice, and also may give different answers with different prompts.  We collected the summaries between June 1 and June 30, 2023, during which no changes were reported to GPT-3.5, though these changes could have occurred unreported.  Note that while the results of this study could change with different prompts or model versions, we view this paper as still valid as a framework for distilling knowledge from large models, and is still reproducible because we release the responses from GPT-3.5 as a separate dataset.  In this way, our procedure is consistent with other research distilling commercial language models, such as~\citet{zhang7llama}.

The subject Java methods are also a key threat to validity because our study results could change with a different set of Java methods.  We reduce the risk by using methods from a large and well-studied dataset, with methods from projects in many domains collected over many years.  Our study used a total of 450 of these methods, randomly selected from the dataset.  The total of 450 is a representative sample: considering a test set population size of 8k methods,~\citet{israel1992determining} sample size recommendations shows a minimum sample size of 381 at +/-5\% precision tolerance.

\subsection{RQ1 Results}

\begin{figure}[]
\centering
\begin{tabular}{ll}
 \includegraphics[width=6cm, height=4cm]{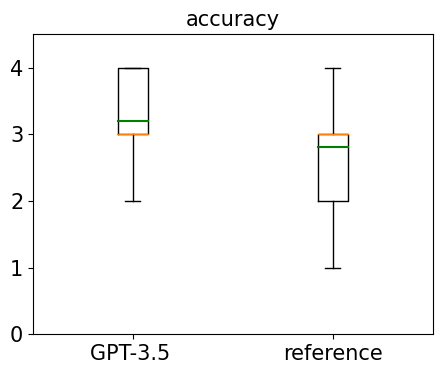} & \includegraphics[width=6cm, height=4cm]{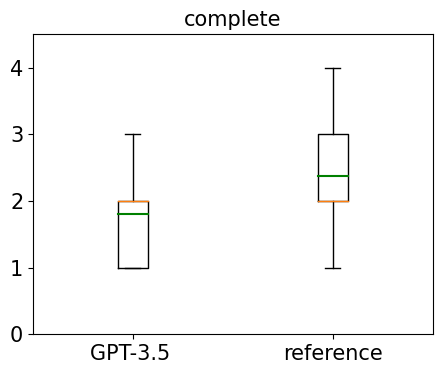} \\
 \includegraphics[width=6cm, height=4cm]{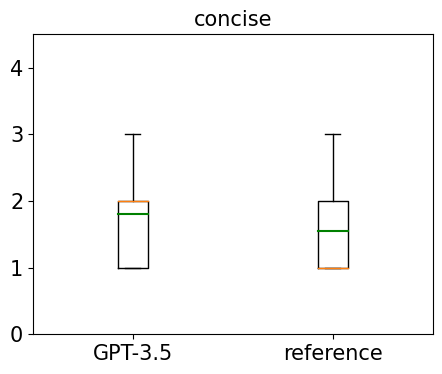}  & \includegraphics[width=6cm, height=4cm]{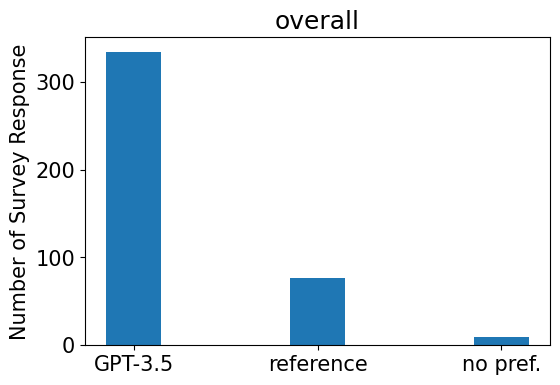}
\end{tabular}
\vspace{1mm}
    {\small
\begin{tabular}{lll|ll|lll}
                              & \multicolumn{2}{c|}{gpt-3.5} & \multicolumn{2}{c|}{reference} &       &       &          \\
                              & med          & mean          & med           & mean           & Zobs  & Zcrit & p        \\ \hhline{~-------} 
\multicolumn{1}{l|}{accurate} & 3            & 3.196         & 3             & 2.811          & 6.816 & 1.645 & $<$0.01  \\
\multicolumn{1}{l|}{complete} & 2            & 1.799         & 2             & 2.368         & 7.670 & 1.645 & $<$0.01  \\
\multicolumn{1}{l|}{concise}  & 2            & 1.808         & 1             & 1.542          & 3.991 & 1.645 & $<$0.01 
\end{tabular}
}
\vspace{1mm}
\caption{Comparison of human reference summaries to GPT-3.5, along the four questions we discuss in Section~\ref{sec:rq1method}, followed by a statistical summary of the RQ1 results in Figure~\ref{fig:rq1results} with Mann-Whitney tests showing statistical significance.}
\label{fig:rq1results}
\vspace{-4mm}
\end{figure}

\begin{figure}[]
\centering
\begin{tabular}{ll}
 \includegraphics[width=6cm, height=4cm]{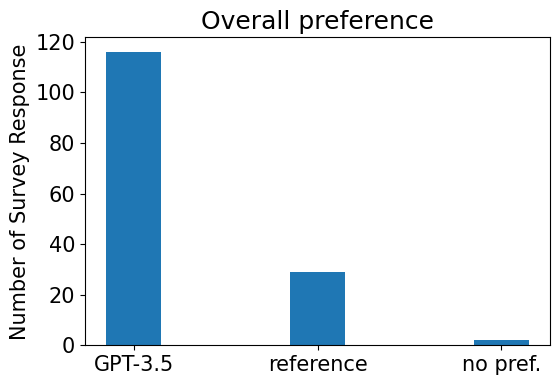} & \includegraphics[width=6cm, height=4cm]{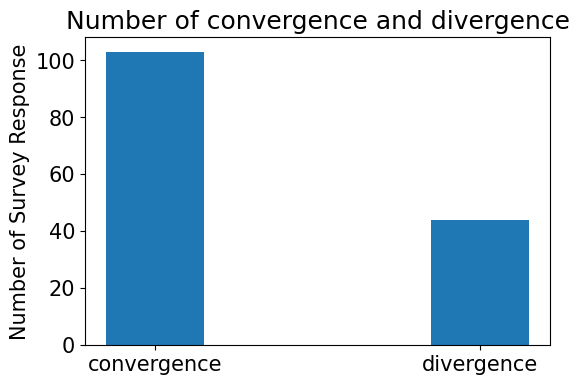}
\end{tabular}

\vspace{1mm}
\caption{Comparison of human reference summaries to GPT-3.5 in terms of overall preference and the number of convergence and divergence functions on the duplicate studies}
\label{fig:rq1resultsduplicate}
\vspace{-4mm}
\end{figure}

\begin{figure}[!b]
\centering
\begin{tabular}{ll}
 \includegraphics[width=6cm, height=4cm]{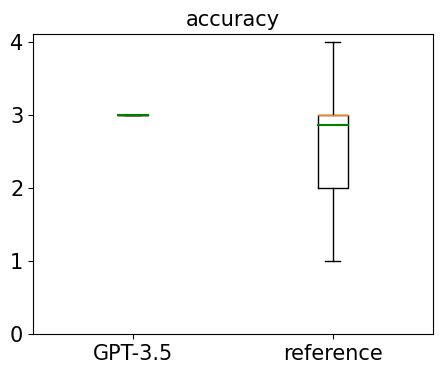} & \includegraphics[width=6cm, height=4cm]{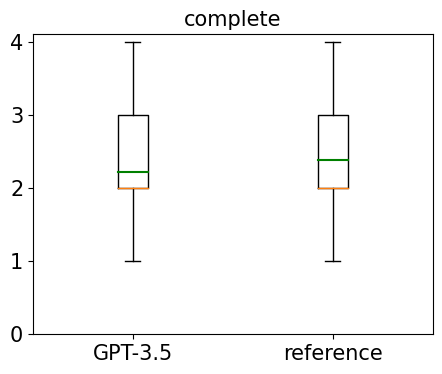} \\
 \includegraphics[width=6cm, height=4cm]{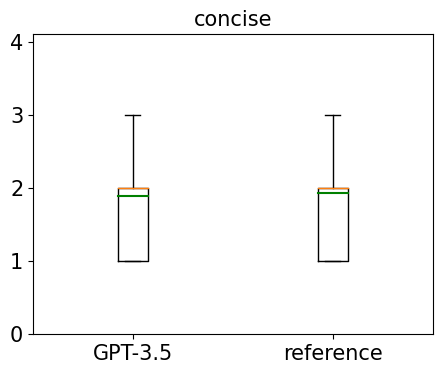}  & \includegraphics[width=6cm, height=4cm]{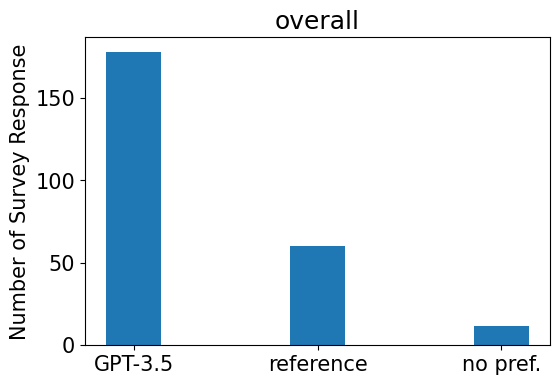}
\end{tabular}
\vspace{1mm}
    {\small
\begin{tabular}{lll|ll|lll}
                              & \multicolumn{2}{c|}{gpt-3.5} & \multicolumn{2}{c|}{reference} &       &       &          \\
                              & med          & mean          & med           & mean           & Zobs  & Zcrit & p        \\ \hhline{~-------} 
\multicolumn{1}{l|}{accurate} &    3      &   2.992    &  3             &    2.860     & -1.232
 & -1.645  &  0.109 \\
\multicolumn{1}{l|}{complete} &    2     &  2.380      &   2         &  2.219     &1.440
 & 1.645 & 0.075 \\
\multicolumn{1}{l|}{concise}  &     2       &  1.890     & 2             &     1.926   & 0.155
 & -1.645 & 0.561
\end{tabular}
}
\vspace{1mm}
\caption{Comparison of summaries generated by models trained with human reference summaries to summaries generated by models trained with GPT-3.5, along the four questions we discuss in Section~\ref{sec:rq1method}, followed by a statistical summary of the RQ1 results in Figure~\ref{fig:rq1resultstrained} with Mann-Whitney tests.}
\label{fig:rq1resultstrained}
\vspace{-4mm}
\end{figure}

We find that GPT-3.5 produces source code summaries that are higher quality than the reference summaries.  The evidence for this finding comes in two key forms, depicted in Figure~\ref{fig:rq1results}.  First, the ratings provided by the human evaluators for accuracy and completeness are better for GPT-3.5 than for the references by a statistically-significant margin.  The mean values for accuracy of GPT-3.5 are higher than the references, with the 90\% of values for GPT-3.5 in the 3-4 range.  In other words, human evaluators marked ``Strongly Agree'' or ``Agree'' to a statement about accuracy for a 90\% of summaries.  For the references, only 74\% were in this range.  Likewise, for completeness, human evaluators rated GPT-3.5 with better scores by a statistically-significant margin (lower is better for completeness and conciseness due to the wording of the survey questions, see Section~\ref{sec:rq1method}).  Second, when comparing summaries, human evaluators preferred GPT-3.5 summaries in 80\% of comparisons, versus 18\% for references and 2\% undecided.  Note that the references were rated more concise that GPT-3.5 summaries, but this result is likely because the references tend to be shorter, with an average of 8.28 words versus 18.40 for GPT-3.5.

Figure \ref{fig:rq1resultsduplicate} depicts the results for the study that we randomly sampled 150 functions from 450 functions in the first study for RQ1. We observe that 79\% of the human evaluation prefers GPT-3.5 summaries, 19\% of the human evaluation prefers human reference, and 2\% of the answers are undecided. The overall results show that GPT-3.5 summaries are the better summary than the human reference and align with the first study.  We observe that  70\% of the functions converge with the first study. We do not calculate the agreement between two studies because \citet{donker1993interpretation, delgado2019cohen} point out that Cohen kappa cannot reflect the agreement in the unbalanced data. All things considered, our results show that GPT-3.5 summary is better than the human reference. 

In addition, we trained the models with 170k human reference proposed by~\citet{leclair2019recommendations} and with the 170k summaries generated by GPT-3.5 for comparison. We use 170k data for comparison because we do not have 2.15m summaries on human reference. We depict the results in Figure \ref{fig:rq1resultstrained}. Overall, the evaluators prefer the summaries generated by the model trained with the GPT-3.5 summaries although the statistical test shows no statistical significance on accuracy, completeness, and conciseness. We find 71\% of the answers prefer the summaries generated by the model trained with GPT-3.5 summaries versus 24\% trained with human reference and 5\% undecided. In terms of accuracy, we observe that the summaries generated by the model trained with GPT-3.5 is slightly higher than the summaries generated by the model trained with human reference. Similarly, evaluators rated the summaries generated by the model trained with the GPT-3.5 summaries more complete than the summaries generated by the model trained with human reference. Although we observe that the results of accuracy and complete only edges to the summaries generated by models trained with GPT-3.5, the summaries generated by the model trained with GPT-3.5 outperforms the summaries generated by the model trained with human reference in overall preference. The possible explanation is that the training data is not enough to mimic the ability of GPT-3.5 for code summarization. For example, compared with ``promotes a Literal numeral to a decimal value by converting it to a Long and creating a new Literal with the converted value and the DECIMAL TYPE'' generated by GPT-3.5 in Table \ref{tab:sumexamples}, the trained model generates ``promotes a given Literal to a decimal type and returns the promoted Literal'', which is less complete and accurate compared with the original GPT-3.5 summary. But, compared with the reference version, this summary is still more informative. Therefore, the evaluator prefers the summary generated by the model trained with GPT-3.5. This result aligns with findings by \citet{roy2021reassessing} that ratings by human evaluators often do not appear statistical significantly different unless the summaries are very qualitatively different due to noise in how people give subjective ratings, yet people may prefer one group of summaries when asked directly. 


This result is surprising considering that human-written references have long been considered the gold standard in evaluating software documentation generation~\citet{leclair2019recommendations, shi2022are}.  Yet consider the examples in Table~\ref{tab:sumexamples}, which are representative of typical summaries from each source.  The human-written references tend to be concise, but lack detail which can lead to confusion (e.g., the meaning of ``reparse the label'' is unclear in the second human-written example in Table~\ref{tab:sumexamples}, but the GPT-3.5 provides detail).  These observations are borne out in the study results where GPT-3.5 is superior in accuracy and completeness, but not conciseness.  They are also supported by evidence in the literature that people often produce low-quality documentation~\citet{aghajani2019software}.  In short, we find that in practice, GPT-3.5 outperforms people when writing software documentation.



\section{Distilling GPT-3.5}
\label{sec:studytwo}

This section discusses the knowledge distillation process and evaluation.  Our objective is to study knowledge distillation of a large language model for code summarization, using the norms for model architecture and datasets prevalent in current literature.  Therefore, we ask the following Research Question:

\begin{description}
\item \textbf{RQ2} How closely do language models mimic GPT-3.5 for code summarization, across different model and dataset sizes?
\end{description}

By ``different model sizes'', we mean models in terms of numbers of parameters.  Much of the current research frontier is focused on Generative Pretrained Transformer (GPT)-like models, with model size considered a major contributor to both model output quality and resource cost~\citep{xu2023small}.  One goal of RQ2 is to help decision-making in balancing model output quality and costs.  It is likely that a ``price break'' will emerge after which the expense of model size will increase faster than output quality gains, and practitioners may wish to choose a model at this break point.  Likewise, by ``different dataset sizes'', we mean size in terms of number of samples.  We collected 2.15m samples from GPT-3.5 in Section~\ref{sec:studyone}, though it is possible that fewer samples are needed.  Additional samples increase training cost, so it may be desirable to use fewer samples.

\subsection{Distillation Process for Decoder-only Language Models}

At a high level, our distillation process is straightforward: we fine-tune a pretrained language model to generate source code summaries, using the summaries generated by GPT-3.5 as training samples.  We use a fine-tuning process proposed by~\cite{su2023language} wherein we use a training prompt of the form:

\vspace{1mm}
\begin{verbatim}
    TDAT: <Java method code>
    COM: <summary of Java method>
\end{verbatim}
\vspace{1mm}

We create training prompts in this form for the entire 2.15m samples we collected from GPT-3.5 in Section~\ref{sec:studyone}.  During fine-tuning, we use the standard autoregressive process in which the model learns to produce each token in the prompt conditioned on all previous tokens.  The model will learn to generate code one token at a time after the \texttt{TDAT}, then learn to generate a summary comment after the \texttt{COM}.  The model will learn to generate summary comments one token at a time, conditioned on the previous tokens in that comment, as well as the Java code prior to the \texttt{COM} token.  In almost all cases, the first word of the summary is an action word such as ``gets'', ``prints'', or ``calculates'', so the model learns to decide this word based on the Java code~\citep{haque2021action}.  The model decides the next word using the action word and the Java code, and continues until it produces an end-of-sequence token to denote the summary's end.

\subsection{Subject Decoder-only Language Models}

We fine-tune two language model architectures as part of our study, using different settings for the model and dataset sizes.  Because the heart of our target model for distillation, GPT-3.5, is a decoder-only Transformer~\citep{brown2020language}, both language model architectures we use are also decoder-only Transformers.  One is a called \texttt{jam} and is a GPT-2-like model that is pretrained using 52m Java methods.  The \texttt{jam} model was released by~\cite{su2023language} as a language model for working with Java code.  The 52m Java methods in the pretraining dataset are searchable for code clones, to limit the possibility of data leakage from test to training set.  The pretraining dataset excludes 8k samples in \texttt{funcom-java-long} test set, which we used in the study in Section~\ref{sec:studyone} and later in this section.  We use the key configuration parameters in Table~\ref{tab:params}, which in a GPT-2-like architecture such as \texttt{jam} result in total network parameter sizes of 350m, 110m, and 38m.

As an alternative, we use \texttt{starcoder}~\citep{li2023starcoder}, which is a 15.5B parameter GPT-2-like model.  The \texttt{starcoder} model is pretrained with ``The Stack'', which is a collection of 6TB of source code in over 350 programming languages.  This model serves as a strong alternative to \texttt{jam} because it represents a state-of-the-art language model, with billions of parameters and an internet-scale pretraining dataset size.  Whereas \texttt{jam} is an inexpensive model with very closely-controlled experimental variables, \texttt{starcoder} is an industrial-strength model with more potential variables due to the much larger and difficult-to-search pretraining data (i.e., there is more risk of data contamination).

\vspace{-4mm}
\begin{table}[!h]
\begin{tabular}{ll|lll|l}
    &                                                  & \multicolumn{3}{c|}{\texttt{jam}} & \texttt{starcoder} \\
$d$ & embedding dimension        &   512   &  768    & 1024  & 6144  \\
$L$ & number of layers           &   4   &    10  & 24    & 40    \\
$h$ & attention heads            &  4    &    8  & 16    & 48    \\
$r$ & learning rate            &  3e-5    &    3e-5  & 3e-5    &   1e-4  \\ 
$e$ & epochs            &  3    &    3  & 3    & 3    \\ 
$o$ & dropouts            &  0.2    &     0.2  &  0.2    &   0.05   \\ 
\hhline{~-----} 
    & total number of parameters & 38m  & 110m & 350m  & 15.5B
\end{tabular}
\vspace{1mm}
\caption{Key model settings and parameters sizes.}
\label{tab:params}
\vspace{-4mm}
\end{table}

\subsection{Subject Encoder-Decoder Language Models}
In addition to decoder-only transformer GPT architecture, we also use encoder-decoder models to distill the knowledge of GPT-3.5 for code summarization. We train \texttt{attendgru} \citep{leclair2019recommendations}, \texttt{transformer}~\citep{ahmad2020transformer}, and \texttt{setransformer}~\citep{li2023setransformer} for 10 epochs on four different datasets. We pick the one with the best accuracy on the validation set as the model for prediction. Table \ref{tab:encdecparams} shows the settings and parameters that we use for encoder-decoder models. We summarize three different models as follows:

\texttt{attendgru} is the model that uses GRU as a backbone with the attention mechanism to form the encoder-decoder architecture. 

\texttt{transformer} uses multi-head attention with the key, query, and value vector for parallelization of the \texttt{attendgru} model. 

\texttt{setransformer} uses transformer as a base with an additional context, abstract syntax tree of the function, as an input and computes the convolution of each input feature.

\vspace{-4mm}
\begin{table}[!h]
\begin{tabular}{l|l|l}
Parameters &     Description   &  Settings     \\
    \hhline{---} 
$d$ & embedding dimension        &   100     \\
$b$ & batch size          &   50       \\
$l$ & learning rate            &  0.001      \\
$s$ & summary vocabulary size             &  10,908      \\
$f$ & functions vocabulary size             &  70k      \\
$t$ & number of tokens for functions            &  50      \\
$c$ & number of tokens for summaries            &  13      \\
\end{tabular}
\vspace{1mm}
\caption{Settings for encoder-decoder models}
\label{tab:encdecparams}
\vspace{-4mm}
\end{table}

\subsection{Dataset Sizes}

We use four dataset sizes during fine-tuning: 2.15m, 1.25m, 620k, and 170k.  The 170k dataset size uses the same Java methods in the training set from \texttt{funcom-java-long} to maintain consistency with previous studies and enable experiments comparing against human-written references (all 170k Java methods in \texttt{funcom-java-long} have human-written summaries, which we remove prior to fine tuning and replace with GPT-3.5-generated summaries).  We then randomly sub-sample the 2.15m dataset of GPT-3.5-generated summaries we created in Section~\ref{sec:studyone} and add these to the 170k, to create 1.25m and 620k datasets.  The 2.15m dataset also contains the 170k samples.  Thus, all dataset sizes contain the same 170k Java methods for comparison, with additional samples added for a maximum size of 2.15m. 

\subsection{Hardware and Software Requirement}
We use NVIDIA RTX A5000 GPU with 24GB VRAM and Intel i9-10900X CPU with
128GB RAM as the hardware to train the models. In addition, we use Pytorch 2.0.1, transformers 4.29.2, and tensorflow 2.12.0 as our software.

\subsection{Evaluation Metrics}

We use two metrics for evaluation: METEOR and USE.  METEOR~\citep{banerjee2005meteor} is a metric that considers the similarity between each word and word overlap for evaluation.  USE~\citep{haque2022semantic} is a metric that encodes the reference and the predicted summary to a fixed-length vector by using universal encoder and computes the similarity scores between two summaries.~\citet{haque2022semantic, roy2021reassessing} point out that METEOR and USE are closer to human preference because these metrics assign partial credits to words instead of treating the importance of the words equally. Therefore, older metrics that only consider word overlap such as BLEU~\citep{papineni2002bleu} are considered as deprecated so we don't report it.

\subsection{RQ2 Results}

\begin{table}[!b]
\begin{tabular}{ll|lll|c|ccc}
         &     & \multicolumn{3}{c|}{\texttt{jam}}  & \multicolumn{1}{c|}{\texttt{starcoder}} & \multicolumn{3}{c}{\texttt{encoder-decoder}} \\
         &      & 38m    & 110m   & 350m  & 15.5B  & attendgru & transformer&  setransformer  \\ \hhline{~--------}
\multirow{4}{*}{\rotatebox{90}{datasets}} & 170k  & 33.88  & 36.71 & 40.73 & 44.8 & 22.39 &  23.17 & 21.33 \\
         & 620k  & 28.29  & 33.98  & 41.57 &     45.59  & 22.88 & 23.59&  22.93   \\
         & 1.25m & 30.19  & 35.58 & 42.63 &  46.38     & 16.64 & 25.10 &  23.18   \\
         & 2.15m & 32.11  & 37.18  & 44.77 &  - &19.53 & 25.38 &  22.45     
\end{tabular}
\vspace{1mm}
\caption{METEOR scores for RQ2.}
\label{tab:rq2meteor}

\vspace{1mm}

\begin{tabular}{ll|lll|c|ccc}
         &     & \multicolumn{3}{c|}{\texttt{jam}}  & \multicolumn{1}{c|}{\texttt{starcoder}} & \multicolumn{3}{c}{\texttt{encoder-decoder}} \\
         &       & 38m    & 110m   & 350m  & 15.5B   & attendgru & transformer&  setransformer  \\ \hhline{~--------}
\multirow{4}{*}{\rotatebox{90}{datasets}} & 170k  & 62.52  & 64.88 & 68.21 &   71.55 & 48.94 & 50.49 & 42.54    \\
         &  620k & 57.78  & 62.84  & 69.24 &     72.16 & 50.17 & 51.19  & 47.27  \\
         & 1.25m &  59.67 & 64.28  & 70.08 &         72.74 & 40.93 & 52.84 & 46.94\\
         & 2.15m & 60.43  & 64.82  & 70.85 & -& 45.14 & 53.33 & 44.83  
\end{tabular}
\vspace{1mm}
\caption{USE scores for RQ2.}
\label{tab:rq2use}

\vspace{1mm}

\begin{tabular}{ll|lll|c|ccc}
         &        & \multicolumn{3}{c|}{\texttt{jam}}  & \multicolumn{1}{c|}{\texttt{starcoder}} & \multicolumn{3}{c}{\texttt{encoder-decoder}} \\
         &       & 38m    & 110m   & 350m  & 15.5B & attendgru & transformer&  setransformer     \\ \hhline{~--------}
\multirow{4}{*}{\rotatebox{90}{datasets}} & 170k  & 0.2  & 0.3 & 1.0 &   28 & 0.05 &0.05 &0.07   \\
         &  620k & 0.5  & 1.5  & 3.5 &  97 & 0.18 &0.18 &0.23        \\
         & 1.25m & 0.9 & 2.5  & 6.5 &  195  & 0.37 & 0.45 & 0.78    \\
         & 2.15m & 2.0  & 4.0  & 10.5 &  -  & 0.67 & 0.63 & 0.78  
\end{tabular}
\vspace{1mm}
\caption{Training time required in hours. The training time for \texttt{jam} and \texttt{starcoder} is the complete finetuning time. The training time for encoder-decoder models is the training time for one epoch.}
\label{tab:rq2time}

\vspace{-3mm}
\end{table}

The automated metrics METEOR and USE indicate a general trend towards better matching of GPT-3.5 as the training dataset and number of model parameters increases, as shown in Tables~\ref{tab:rq2meteor} and~\ref{tab:rq2use}.  The METEOR score for the 170k dataset using the 38m parameter \texttt{jam} model is 33.88, which rises to 40.73 for the 350m parameter \texttt{jam} model and 44.8 for \texttt{starcoder}.  Likewise, the 40.73 score for the 350m \texttt{jam} model rises to 44.77 as the dataset increases from 170k to 2.15m samples.  The USE scores show the same pattern, with the 350m \texttt{jam} model ranging between 68.21 and 70.85 as the dataset size increases.  This pattern is not surprising given that larger model and dataset sizes are widely regarded as resulting in better automated scores in many domains~\citep{schaeffer2023emergent}, and are consistent with a view that the models are able to learn to mimic at least a portion of GPT-3.5's ability to summarize code.

A ``price break'' occurs favoring the use of the 350m \texttt{jam} model.  Table~\ref{tab:rq2time} shows that \texttt{starcoder} is 28 times more expensive than the 350m \texttt{jam} model, yet reaches only 10\% higher METEOR and 5\% higher USE scores.  The \texttt{starcoder} model was not feasible to fine tune for the 2.15m dataset due to an estimated 14 days cost requirement, while previous metric increases were low (less than 1\% difference in USE between 1.25m and 620k datasets, for example).  In contrast, the 350m \texttt{jam} model cost is only slightly higher than smaller models, and can be operated on a single 16GB GPU~\citep{su2023language}.  Note that while it is tempting to write off training (or even inference) costs as sunk costs, in fact model expense is a key engineering detail affecting model deployment and cost/benefit analyses in industrial products~\citep{bender2021danger}.  Overall, the 350m \texttt{jam} model provides a balance between cost and performance.

We find that the performance of some models such as 38m and 110m parameter \texttt{jam} does not increase as the data size increases. For example, in the 38m parameter \texttt{jam}, 170k data size outperforms any other datasets that are larger than 170k.~\citet{perez2021much} also observed the similar phenomenon that the performance of certain models do not always increase as the data size increases in the syntactic generation task. The possible explanation is that models learn the trivial features when the models are not large enough for the certain size of dataset~\citep{chang2023language}. Also, the smaller models saturate faster than the larger models~\citep{zhai2022scaling}. This can further show that the 350m parameter \texttt{jam} is small enough to operate on 16GB GPU~\citep{su2023language}, but large enough to mimic the portion of GPT-3.5's ability for code summarization.

In terms of the encoder-decoder language models, we find that the \texttt{attendgru} and \texttt{setransformer} follow the trend of reverse performance when they reach certain amount of data size. For example, \texttt{attendgru} has the best performance at 620k data size instead of 2.15m data size. This is because the models are much smaller than our pretrained models. Compared with the pretrained 38m and 110m parameter \texttt{jam} models, \texttt{attendgru} reaches the plateau at 620k data size instead of 170k data size because we train the model from scratch. Although we observe that performance of \texttt{transformer} increases as the data size increases, the improvement of \texttt{transformer} is relative small compared with 350m parameter \texttt{jam} (4.8\% improvement on 350m parameter \texttt{jam} versus 1.1\% on \texttt{transformer} between 2.15m and 1.25m dataset, for example). All things considered, we show that 350m parameter \texttt{jam} is the better model for the knowledge distillation for code summarization.   

\section{GPT-3.5 and Distilled Summaries Comparison}
\label{sec:studythree}

This section discusses our comparison with human experts of source code summaries generated by GPT-3.5 to summaries generated by the 350m parameter version of the \texttt{jam} model trained with the 2.15m example dataset.  The purpose of this comparison is to measure how closely the \texttt{jam} model replicates GPT-3.5 for the task of code summarization.  We chose to compare the 350m parameter version of the \texttt{jam} model instead of the various alternatives in Section~\ref{sec:studytwo} because the 350m \texttt{jam} model achieves a balance between performance and affordability.  This model is within 10\% performance of \texttt{starcoder} in terms of METEOR and USE scores (e.g., 44.8 versus 40.73 METEOR), while also requiring only a single 16GB GPU a relatively short amount of time.  The \texttt{starcoder} model does have higher performance, but at much higher cost.  Meanwhile, the 350m parameter \texttt{jam} model achieves a greater-than 10\% improvement in METEOR and USE scores over the 110m parameter model, at relatively low cost in practice.  The 350m parameter \texttt{jam} model is a balance between cost and performance.  To compare \texttt{jam} to GPT-3.5, we ask the following RQ:

\begin{description}
\item \textbf{RQ3} How closely does the distilled model mimic GPT-3.5 for code summarization, as measured by human experts?
\end{description}

Our rationale for asking this question is that while the study in Section~\ref{sec:studytwo} measures how well different models mimic GPT-3.5 in terms of automated metrics, these automated metrics are known to diverge from human expert opinion at times~\citep{novikova2017why}.  To fill this potential gap, we also perform a study with human experts to compare the models based on human preference.  Our research method to answer RQ3 is identical to how we answer RQ1, except that we change the source of the samples and recruit a new group of participants.  We use the same survey pages and ask the same questions.  As with RQ1, we do not mark the summaries as coming from any particular model, to avoid demand characteristic bias~\citep{dell2012yours}.  We recruit 15 new participants via the Prolific platform using the same criteria as for RQ1. Also, we randomly select 150 functions from 450 functions that we use for the first study to answer RQ3 and recruit five different participants to evaluate the summary with the same criteria to reduce bias as in RQ1.
\begin{figure}[]
\centering
\begin{tabular}{ll}
 \includegraphics[width=6cm, height=4cm]{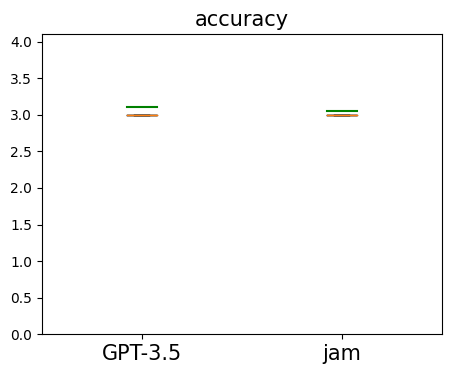} & \includegraphics[width=6cm, height=4cm]{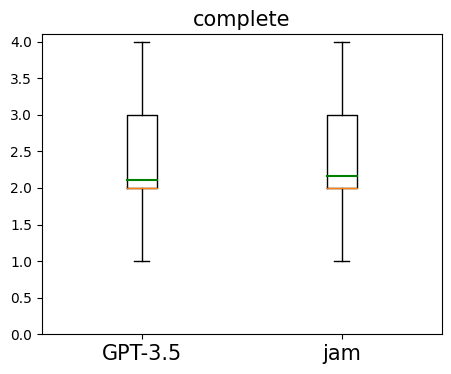} \\
 \includegraphics[width=6cm, height=4cm]{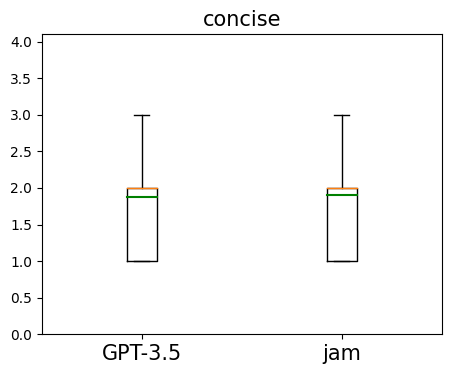}  & \includegraphics[width=6cm, height=4cm]{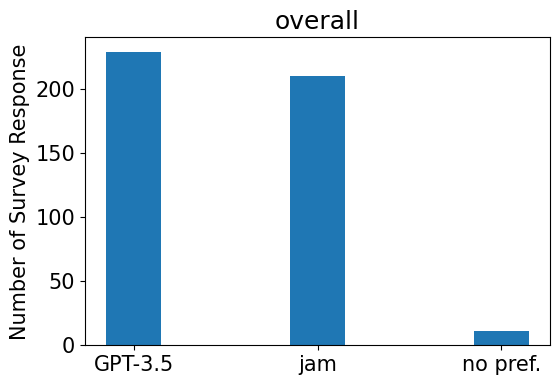}
\end{tabular}
\vspace{1mm}
    {\small
\begin{tabular}{lll|ll|lll}
                              & \multicolumn{2}{c|}{gpt-3.5} & \multicolumn{2}{c|}{jam}       &       &       &          \\
                              & med          & mean          & med           & mean           & Zobs  & Zcrit & p        \\ \hhline{~-------} 
\multicolumn{1}{l|}{accurate} & 3            & 3.112         & 3             & 3.051          & 1.009 & 1.645 & 0.27     \\
\multicolumn{1}{l|}{complete} & 2            & 2.107         & 2             & 2.162          & 0.600 & 1.645 & 0.73     \\
\multicolumn{1}{l|}{concise}  & 2            & 1.902         & 2             & 1.876          & 0.244 & 1.645 & 0.40 
\end{tabular}
}
\vspace{1mm}
\caption{Comparison of \texttt{jam} to GPT-3.5 followed by a statistical summary of the RQ3 results with Mann-Whitney tests showing statistical significance.}
\label{fig:rq3results}
\vspace{-4mm}
\end{figure}

\begin{figure}[]
\centering
\begin{tabular}{ll}
 \includegraphics[width=6cm, height=4cm]{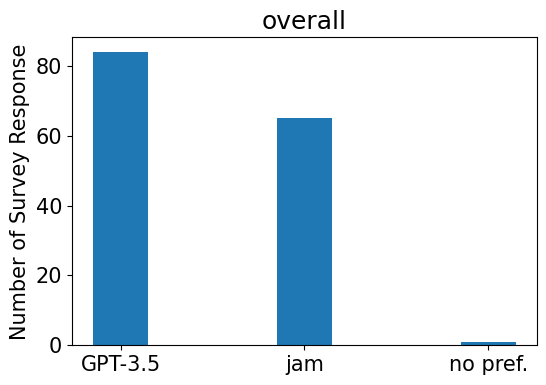} & \includegraphics[width=6cm, height=4cm]{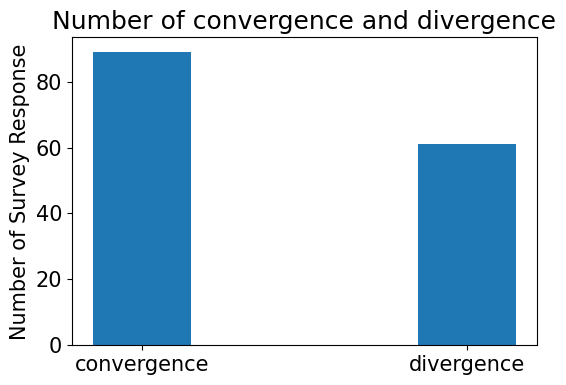}
\end{tabular}

\vspace{1mm}
\caption{Comparison of \texttt{jam} to GPT-3.5 in terms of overall preference and the number of convergence and divergence functions on the duplicate studies}
\label{fig:rq2resultsduplicate}
\vspace{-4mm}
\end{figure}

Figure~\ref{fig:rq3results} depicts the results.  In short, we do not observe a statistically-significant difference between the 350m parameter \texttt{jam} model and GPT-3.5 in terms of accuracy, completeness, or conciseness.  The mean value for accuracy and completeness for GPT-3.5 is better than \texttt{jam}, though these differences are not statistically-significant.  Likewise, the mean value for conciseness is slightly better for \texttt{jam}, but is also not significant.  In terms of overall preference, participants preferred summaries from GPT-3.5 52\% of the time versus \texttt{jam} 46\% of the time (2\% undecided).  Taken together, these results support a conclusion that \texttt{jam} reproduces GPT-3.5's source code summarization, though with a slight edge to GPT-3.5 in terms of overall preference.

Figure \ref{fig:rq2resultsduplicate} shows the results for the study for the 150 functions that we randomly selected from 450 functions in the first study for RQ3. We find participants prefer GPT-3.5 56\% of the time versus the 350m parameter \texttt{jam} model 43\% of the time and 1\% undecided. This result aligns with the first study that the overall preference edges to GPT-3.5. In terms of the convergence rate, we find only 59\% of the answers converge with the first study, which is lower than the results in RQ1. This is expected because the 350m parameter \texttt{jam} model aims to reproduce the ability of the GPT-3.5 for code summarization. This can further show the 350m parameter \texttt{jam} model's ability for reproduction of GPT-3.5 for code summarization although the overall results still show that GPT-3.5 is slightly better.

Yet these results do not mean that the summaries from \texttt{jam} and GPT-3.5 are identical.  Consider Table~\ref{tab:sumexamples2}.  Participants favored the first two samples from GPT-3.5 and the second two from \texttt{jam}.  While we cannot include full code due to space limitations, these samples illustrate how the models provide broadly similar content, though with different surface realization of the text or different decisions about what information to condense.  As a general observation, we note that GPT-3.5 tends to be more verbose, though whether this verbosity is perferred is unclear.  For example, the second example shows how GPT-3.5 wrote ``reads the next byte from a buffer and returns the corresponding character'' whereas \texttt{jam} wrote ``reads a character from a buffer'' for the same information.  The \texttt{jam} summary is more concise but less complete, but the GPT-3.5 summary was preferred.  Likewise, the third sample shows how \texttt{jam} writes ``with specified paramters'' rather than listing them like GPT-3.5 does, though in this example \texttt{jam} is preferred.

\begin{table}[]
\begin{tabular}{c|l|p{10cm}}
preferred & model & summary \\ \hline
$\times$  & GPT-3.5 &  recycles a view for a given index, checking whether the index is within the range of items and whether the view needs to be added to the empty items or items list
\\
          & jam     & receives a View object and an index, and if the index is out of range, it adds the View to the wheel and sets the emptyItems flag to true
 \\ \hhline{~--}
$\times$  & GPT-3.5 &  reads the next byte from a buffer and returns the corresponding character, while throwing an exception if there is not enough data available
\\
          & jam     & reads a character from a buffer and throws an exception if there is not enough data available
 \\ \hhline{~--}
          & GPT-3.5 & inserts a new item into the database with the given attributes such as user ID, category, description, start/end dates, minimum bid, and name
 \\
$\times$  & jam     & inserts a new item into a database table with specified parameters
 \\ \hhline{~--}
          & GPT-3.5 &  searches for a specific Physical2DObject in a Physical2DEnvironment based on its id and returns it
\\
$\times$  & jam     & retrieves a Physical2DObject from an ArrayList based on a given ID
\end{tabular}
\vspace{1mm}
\caption{Examples of summaries generated by GPT-3.5 and \texttt{jam}.  We show two in which each model was preferred in the human study.}
\label{tab:sumexamples2}
\vspace{-2mm}
\end{table}

\section{Discussion/Conclusion}

This paper moves the state-of-the-art forward in three key ways:

\begin{enumerate}
    \item We present a study comparing source code summaries generated by GPT-3.5 to summaries provided as references with the code itself.  We found that human readers preferred the summaries generated by GPT-3.5 by a significant margin.  This result has two important implications.  First, AI-based models are likely to be useful tools for augmenting or even replacing documentation written by people, which supports a vision of researchers for on-demand developer documentation~\citep{robillard2017demand}.  Second, the research community may reconsider using human-written references as the gold standard for training and evaluating source code summarization tools.  AI-based models may be superior at times.
    \vspace{1mm}
    \item We present a study in which we distilled GPT-3.5's code summarization abilities into several smaller models and thoroughly examine the exact model and data size for knowledge distillation on code summarization.  Some of these models are several orders of magnitude smaller than GPT-3.5, and yet are able to achieve comparable results in a large portion of instances when measured by the automated metrics METEOR and USE.  Specifically, we observe that 350m jam model with 2.15m data is a tradeoff. A key advantage to these models is that they can be run locally, with tractable costs for many consumers (a single 16GB consumer GPU, for instance).  Local model execution means local custody of data.  With our distilled model, it is possible to replicate much of the benefit of a large model for code summarization, without losing control of ones sensitive data and source code.
    \vspace{1mm}
    \item We present a study comparing GPT-3.5 to a distilled model (350m parameter \texttt{jam}) with human experts.  This study shows how \texttt{jam} is able to reproduce GPT-3.5 for code summarization.  We did not observe a statistically-significant difference between the summaries from the two models in terms of accuracy, completeness, or conciseness.  We did observe a slight preference favoring GPT-3.5 in direct comparisons by participants (52\% GPT-3.5, 46\% \texttt{jam}, 2\% undecided).  Overall, these results support a conclusion that an inexpensive model such as 350m \texttt{jam} can replicate a very large model such as GPT-3.5 for the task of code summarization when provided sufficient examples.
\end{enumerate}

Finally, we release all data and code for our studies and approach via an online appendix.  We encourage reproducibility of our results, as well as access to the technology via our implementation in Data and Code Availability Section.

\section*{Funding}
This work is supported in part by NSF CCF-2100035 and CCF-2211428. Any opinions, findings, and conclusions expressed herein are the authors and do not necessarily reflect those of the sponsors.
\section*{Contributions}
Chia-Yi Su did the experiments and implemented the code for finetuning tasks. Collin McMillan wrote the manuscript, built the infrastructure and generated the dataset for experiments. 

\section*{Conflict of Interest}

The authors declare that they have no competing interests.

\section*{Data Availability}

All of the datasets and models are in our APCL Hugginface repository,
{\url{https://huggingface.co/datasets/apcl/Jam-CGPT}} and 
{\url{https://huggingface.co/apcl/Jam-CGPT}}

\section*{Code Availability}\label{sec:codeavalibility}

We release our code for experiments in our APCL Github repository, 
{\url{https://github.com/apcl-research/Jam-CGPT}} 



\bibliography{sn-bibliography}
\end{document}